\documentclass{article}

\usepackage{tablefootnote}

\usepackage{natbib}

\usepackage{graphicx}

\usepackage{authblk}
\usepackage{endnotes}
\usepackage{tabularx,caption}
\usepackage{float}
\usepackage{pgfplots}
\usepackage[eulergreek]{sansmath}        
\pgfplotsset{
	compat=newest, 
	small, every non boxed x axis/.append style={x axis line style=-},
	every non boxed y axis/.append style={y axis line style=-}
}

\pgfplotsset{                            
	tick label style = {font=\sansmath\sffamily},
	every axis label = {font=\sansmath\sffamily},
	label style = {font=\sansmath\sffamily}
}

\date{}

\begin{document}

\title{Digital Humanities\\ Readiness Assessment Framework:\\ DHuRAF}
\author{Hossein Hassani \thanks{Department of Computer Science and Engineering, University of Kurdistan Hewl\^er} \thanks{hosseinh@ukh.edu.krd} \and Emir Turajli\'c \thanks{e.turajlic.sa@gmail.com} \and Kemal Taljanovi\'c \thanks{ktaljanovic@yahoo.com}}

\maketitle

\begin{abstract}
	This research suggests a framework, Digital Humanities Readiness Assessment Framework (\textbf{DHuRAF}), to assess the maturity level of the required infrastructure for Digital Humanities studies (\textbf{DH}) in different communities. We use a similar approach to the Basic Language Resource Kit (BLARK) in developing the suggested framework. DH as a fairly new field, which has emerged at an intersection of digital technologies and humanities, currently has no framework based on which one could assess the status of the essential elements required for conducting research in a specific language or community. DH offers new research opportunities and challenges in the humanities, computer science and its relevant technologies, hence such a framework could provide a starting point for educational strategists, researchers, and software developers to understand the prerequisites for their tasks and to have a statistical base for their decisions and plans. The suggested framework has been applied in the context of Kurdish DH, considering Kurdish as a less-resourced language. We have also applied the method to the Gaelic language in the Scottish community. Although the research has focused on less-resourced and minority languages, it concludes that DHuRAF has the potential to be generalized in a variety of different contexts. Furthermore, despite significant reliance on Natural Language Processing (NLP) and computational utilities, the research showed that DH could also be used as an essential resource pool to leverage the NLP study of less-resourced and minority languages.


\end{abstract}
\let\footnote=\endnote
\setlength{\parskip}{1em}
\section{Introduction}
\label{intro}

Currently, there is no framework or an accounting mean that could be used to capture the state of DH studies in a specific context. Such framework can play a significant role in identifying the requirements for panning to conduct infrastructural projects that could advance DH studies. It can also help in finding gaps in different sections of DH in a particular environment.

This paper suggests a framework that we have called it Digital Humanities Readiness Framework (DHuRAF). We use a similar approach to Basic Language Resource Kit (BLARK) in developing the suggested framework. The Basic Language Resource Kit (BLARK) is an informative framework that shows the capability and maturity level of a language with regard to Natural Language Processing (NLP) and Computational Linguistics (CL)~\citep{krauwer1998elsnet,krauwer2003basic}. Similarly, DHuRAF identifies a number of parameters such as required tools, technology, and educational support which are considered to be related to the situation of DH in a specific community, region, or country whereby. Using these parameters, DHuRAF shows the current situation and the possible gaps between that current situation and the expected status that must be addressed. DHuRAF has also been designed in a way that is generalizable. That is, it could be used in different contexts of DH assessment, though in this research we have mainly targeted communities that use minority languages or languages that are digitally less-developed. As an experiment, we have applied the framework in the Kurdish studies context and assessed the result to reach a conclusion about DH in the Kurdish context.

Digital Humanities (DH) is an area of research born through intersection of humanities and computer science that combines linguistics, literature, philosophy, history, music and art with information retrieval, data and text mining, categorization, clustering and data visualization. The emergence of DH is believed to be an inevitable aspect of humanities studies in the coming years~\citep{terras2016decade}. This emergence is happening in such a way which it even extends beyond the scopes of humanities field and embraces quantitative methods and techniques from social and natural sciences as well. As a wide encompassing field, not only providing a comprehensive definition for DH but also tracing its historical developments are considered as topics of debate for many scholars.

In other words, DH ``is a new set of practices, using new sets of technologies, to address research problems of the discipline'' of humanities~\citep{borgman2009digital}. According to some scholars, DH finds its roots in humanities computing, which dates back to 1949~\citep{companion2DH}. However, it still is considered as ``a young field in rapid development''~\citep{KCL1} in which `` [f]rom the outside looking in, [its professional researchers] seem amateur''~\citep{terras2012present}. From another perspective, \cite{svensson2012envisioning} believes that ``digital humanities is serving as a means to advocate and rethink the Humanities''.

There are also thoughtful and philosophical discussions about the situation of DH as a standalone science~\citep{thaller2012controversies,thaller2012digital}. According to~\cite{burdick2012digitalhumanities} DH neither just focuses on the digital culture nor only follows the traditional humanities. Instead, it establishes itself as a distinctive interdisciplinary discipline through ``the opportunities and challenges'' that the combination of the humanities concept and digital artifacts provides~\citep{burdick2012digitalhumanities}. From this perspective, applying the traditional humanities to the digital context is one of the main ingredients of DH. 

Regardless of the discussions about the definition of DH and its origins, when one looks at the overall picture, one can realize that the efforts in advancing Digital Humanities are not only emerging but also significantly changing the overall scene of humanities studies. However, this, as the situation suggests, is only happening with regard to those societies and environment where they already have an established or nearly-established infrastructure and required technology. That is, the tools, technology, and material necessary for research projects in DH are ready and are supported by educational activities and institutions. The countries and regions that lack the basic requirements should first plan to build the necessary foundations.

DH significantly relies on computing tools and techniques~\citep{puschmann2015digital}. In their two important companions to DH and Digital Library Studies, \cite{schreibman2008companion} have provided a thorough figure on the relations between Digital Humanities and computing from different perspectives. Some researchers have addressed particular areas of these relations. For instance, ~\cite{Marsden01112007} has emphasized the importance of software tools, especially audiovisual tools, and their role in Digital Humanities research. On the other hand, concerns have been shown about the ignorance of traditional humanities research methods in using the digital resource and computational approaches~\citep{rohle2012digital}.

Furthermore, several large projects have been undertaken in developed countries in order to bring the efforts of different researchers into an interdisciplinary environment to advance DH. PioNEER is a significant sample of this kind of projects~\citep{Springer1}. Another sample is the CULTURA project, which has developed an extensive powerful platform, based on a rigorous architecture, which allows DH research, particularly on the cultural heritage, to emerge and flourish~\citep{Springer2}.

The rest of this article is organized into three sections. Section one presents the proposed framework. Section two uses the framework in the context of Kurdish language. The third section shows the application of the framework in the context of Gaelic language. Finally, the last section summarizes the findings and provides the conclusions.

\section{Digital Humanities Readiness Assessment\\ Framework (\textbf{DHuRAF})}
\label{ssec:DHMA}

To be able to study Humanities \textit {digitally}, the target environment must be \textit {computationally} ready~\citep{berry2011computational,berry2012introduction}. DH are highly dependent on information technology. That is, DH scholarship requires a cyber-infrastructure, i.e. ``sharing advanced computing infrastructure, training in advanced technologies for humanities research, and developing repositories for digital collections~\citep{zorich2008survey}.'' Surveys also show the strong correlation between academic studies and research in DH with the availability of computational technologies in many countries~\citep{russell2014geographical}.

Our search for an established method or framework that addresses the ``DH~readiness'' or ``DH maturity level'' that could be used as a base for assessing the fundamental requirements that should be considered as the necessary prerequisites for scholarship studies of DH to be emerged was not fruitful. Variety of academic studies have addressed the DH situation from different angles of view (see~\citep{companion2DH,schreibman2008companion,zorich2008survey}), however, we did not come across a particular framework that could be applied to different cases or could be generalized. By ``framework'' we mean a set of established parameters and measures that are able to, preferably quantitatively, determine the state of a certain subject, which in our case is DH.

To put it simply, we need a framework, which by using it we are able to figure the status of DH in the context of specific, for instance Kurdish, studies. The figure should tell us, for example, that the DH studies and required repositories \textit{does not exist}, \textit{premature}, \textit{in its preliminary stage}, or \textit{mature/fully functional}. As a result, we developed a framework that is assessing the readiness status of a specific community/language/region for DH. We have called this framework Digital Humanities Readiness Assessment Framework (\textbf{DHuRAF}). It uses a combination of qualitative and quantitative parameters that together are able to depict a detail figure of the DH status in the target community or environment.

In developing DHuRAF we have followed a method that has been used in BLARK (Basic Language Resource Kit) in Natural Language Processing (NLP) and Computational linguistics (CL). BLARK is an informative framework and also resource kit that shows the capability and maturity level of a language with regard to NLP and CL ~\citep{krauwer2003basic}. Like BLARK, DHuRAF identifies some parameters based on which the situation of DH in a specific context would be quantified. DHuRAF has been designed in a way that is generalizable. That is, it could be used in different contexts of DH assessment, though in this research we have mainly targeted communities that use minority languages or languages that are digitally less-developed. Below the framework is explained. 

\subsection{The Architecture of DHuRAF }

\begin{figure}[h]
	\scalebox{.94}{\input{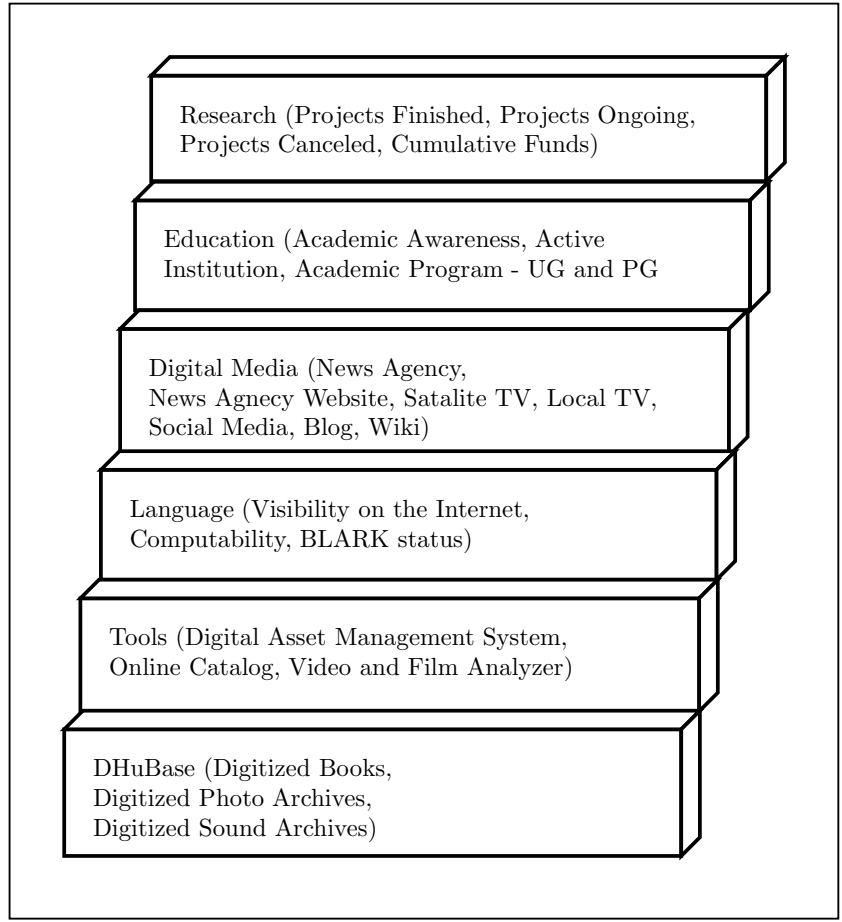}}
	\caption{DHuRAF Architecture - This architecture shows the 6 basic components of DHuRAF.}
	\label{fig:DHuRAF}
\end{figure}

The architecture of DHuRAF consists of several main components. Figure \ref{fig:DHuRAF} shows a general view of this architecture. The architecture and its components are described in the following sections.

The general structure of DHuRAF is based of the followings:
\begin{itemize}
\item DHuRAF consists of 6 components/sections. 
\item Each component includes several items. 
\item Each item has two measurements: \textbf{importance} and \textbf{availability} which are explained below. 
\subsubsection{Imporatnce} \textit{Importance} indicates how important this item is according to the DH academics and/or professionals. It is measured using \textbf{+} signs. The number of signs, from 1 to 3, shows the importance degree. That is, the more pluses indicates higher importance. Leaving this measure unfilled indicates an undecided situation. This might be because no enough information could be obtained or the academics and/or professionals have not been able to make a decision at this stage. 
\subsubsection{Availability} \textit{Availability} indicates how available the item is based on the collected data. It measures the item either quantitatively or qualitatively. 
\end{itemize}

It seems that there is a correlation between \textit{importance} and \textit{availability}, that is, an increase in one should increase the other. However, DHuRAF does not consider such correlation and encourages that these indicators should be evaluated independently. This ensures an unbiased measurement without incorporating any presumption. Below the measures and their descriptions are provided. 

\begin{itemize}
\item \textbf{DHuBase}- Digital Humanity Base has been borrowed from computer database concepts. This section summarizes the data and material that are ready for the DH processes. In order to evaluate the quality of the material which form the contents of this database, we need to use supportive procedures and guidelines. \cite{gonccalves2007good}, for example, proposed a framework to evaluate a digital library. The framework utilizes a wide range of indicators by which a digital library is assessed. The framework assumes that the quality assessment of digital libraries could be automated without raising a significant concern about the quality properties which might be the source of concerns with regards to physical libraries. Additionally, to decide whether to consider a digital resource to be qualified to be added to this database or not depends on different parameters several of which have been suggested by~\citep{hasan2009automatic} in a model that is used to qualify wiki resources. Furthermore, \citep{stvilia2007framework} introduced a model that evaluates the quality of information.
\begin{itemize}
\item Digitized Books- The number or the amount of digitized books. 
\item Digitized Photo Archives- The number or the amount of digitized photo archives. 
\item Digitized Sound Archives- The number or the amount of digitized sound archives. 
\end{itemize}
\item \textbf{Tools}- The tools that are used in DH. It should not be confused with the tools which are used in NLP.
\begin{itemize}
\item Digital Asset Management System- Is there any system available that could manage digital assets? There might be systems that are not language and/or cultural dependent. However, the focus here is on those that are language and/or cultural dependent.   
\item Online Catalog- Is there any online catalog that is specialized for the target community/language/culture? 
\item Video and Film Analyzer- Although highly dependent on languages computability, there might be some tools available for analyzing videos and films specialized for the specific community. For example, there are films (videos) with subtitles (captions) in the target language that can be processed and used. This is possible for these systems to have a blurred overlap with language tools in NLP.
\end{itemize}
\item \textbf{Language}- Languages play a central role in DH. This section summarizes the language capability to be utilized by the DH processes.
\begin{itemize}
\item Visibility on the Internet- How visible the language is. This item is a qualification that can be set by quantification. For example, how many pages would be found for some random keywords? 
\item Computability- Is the language computable? Are there language tools that make the computing process efficient? How reach the LT (language technology) is? 
\item BLARK Status- Is there any BLARK (or similar to BLARK) for the language? If the answer is yes then this indicator is also considered an indicator for the preceding item.
\end{itemize}
\item \textbf{Digital Media}- Digital media is one of the main sources of DH. This section summarizes the status of digital media for the target environment.
\begin{itemize}
\item News Agency- News agencies are the sources of mass digital context generation. Their number is considered as an indicator of readiness for DH processes.   
\item News Agency Website- News agencies website are also the sources of mass digital context generation. Their number is considered as an indicator of readiness for DH processes.   
\item Satellite TV- TV channels are the sources of mass digital graphics. Their number is considered as an indicator of readiness for DH processes.   
\item Satellite TV Website- TV Websites are the sources of mass digital context generation. Their number is considered as an indicator of readiness for DH processes.   
\item Local TV- Local TVs can be a window to cultural aspects that could even if produced in analog, the could be converted to digital to be used as a resource.   
\item Social Media (Twitter, Facebook, ...)- The existence of social media and their magnitude. 
\item Blog- The number or the amount of blogs. 
\item Wiki- The number or the amount of wikis. 
\end{itemize}
\item \textbf{Education}- The Higher education status is considered as an indicator of the maturity of DH in a target community.
\begin{itemize}
\item Academic Awareness- This indicator can be inferred trough either qualitative or quantitative approach. The following items can be considered as quantitative measures. If these measures are not promising then a qualitative approach, for example, by conducting interviews with the academics in the target community or interested in studying the target community in the context of DH can be used. Also other quantitative methods, for example, through designing and using questionnaires can be applicable.  
\item Active Institution- The number of active institutions in DH. 
\item Academic Program-UG- The number of programs related to DH at undergraduate level. 
\item Academic Program-Master- The number of programs related to DH at master level. 
\item Academic Program-PhD- The number of programs related to DH at doctorate level. 
\end{itemize}
\item \textbf{Research}- The depth and breadth of DH related research projects in the target community or related to the target community.
\begin{itemize}
\item Projects-Finished- The number of accomplished projects. 
\item Projects-Ongoing- The number or ongoing projects. 
\item Projects-Canceled- The number or the projects canceled. 
\item Cumulative Fund- The cumulative amount that has been spent in DH research up to date. 
\end{itemize}
\end{itemize}

If an exact number is not available, the amount can be enumerated as 3: vast, 2: considerable, 1: available, 0: not available.

The following section provides suggestion on how to interpret the results of prepared DHuRAF Indicators.

\subsection{DHuRAF Indicators Interpretation}
The DHuRAF Indicators provide a summarized yet detail view of the DH situation in a target community. Nevertheless, one might be interested in categorization of the overall outcome. For this, we suggest the following categories:

\begin{itemize}
\item \textbf{Void}- Not ready for DH studies. There is no or very rare indication of DH activities.
\item \textbf{Infancy}- DH is in its infancy. There are some indications or few evidence of DH activities. 
\item \textbf{Premature}- DH is premature. There are indications of average numbers of evidence in nearly all areas of DH activities.
\item \textbf{Mature}- DH is mature. There are indications of large numbers of evidence in all areas of DH activities.
\item \textbf{Flourished}- DH is flourished. There are indications of vast numbers of evidence in all areas of DH activities.
\end{itemize}   

The above categories are general indicators of the DH status in the target community. They should be used alongside the detailed information which accompanies the framework. In fact, this is this detailed information that helps researchers and developers to conduct the necessary projects whereby could help the DH status in the target community to be improved. In the next section, the DHuRAF will be applied in Kurdish case in which we discuss the interpretation of the indicated category. 

\section{Applying DHuRAF: Kurdish Case}

To investigate the efficiency of DHuRAF, it has been applied for Kurdish. The following sections provide the results of applying DHuRAF and the analysis of the results.

\subsection{Findings}

The findings have been documented according to the DHuRAF sections and have been summarized using the DHuRAF Indicators. This section explains how the findings have been obtained and shows the DHuRAF Indicators table.  

\subsubsection{DHuBase}

For DHuBase, we basically used our priori knowledge about organizations and institutions which are active in Kurdish studies. Below some examples have been presented. However, there are other institutions which perform important activities in this regard that have not been listed here. This exploratory approach, with a qualitative viewpoint in mind, aimed in finding evidence that could be taken into account as one of the items of DHuBase section. This was regardless of the fact that the evidence were or were not considered as a Digital Humanities activity from the providers point of view.

\begin{itemize}
\item Kurdish Institute of Paris - It has organized a library, which is known as BNK ({La biblioth\'eque num\'erique kurde}\footnote{The Kurdish Digital Library}), that includes about 730 Kurdish items, some of which are available in pdf format~\citep{BNKIKP}.
\item SARA Publication - SARA manages a website, which provides a collection of digital items related to Kurdish culture. This collection includes more than 300 digital items of different topics~\citep{SARA1}. It has also collected resources on the Kurdish literature~\citep{SARA2}. Furthermore, the website provides a useful link page that presents a rich reference to different Kurdish related sources on the Internet~\citep{SARA3}. 
\item Kurdish Heritage Foundation of America - This is a combination of digital and non-digital resources, which was donated by ``The Foundation for Kurdish Library \& Museum'' to the Binghamton University. It is a rich resource of diverse kinds of about 1000 digitized cultural Kurdish items~\citep{saeedpourKLM}.
\item Kurdipedia - This is yet another rich library of Kurdish literature, some of which available in pdf format. An interesting fact is that the site includes some completely rewritten resources in a script which is different than the original script of the resource. For example, a poetry in Sorani dialect, originally in Arabic/Persian script, is available, which has been rewritten in Latin script~\citep{kurdipedia1}. 
\item Ferheng.org - Ferheng website provides an online dictionary and forum that is open for Kurdish related discussion~\citep{ferheng1}.
\item Kurdish Heritage Institute - An NGO which has been established in 2003 with the aim of creating an archive for cultural heritage \citep{khi}.
\item Kurdish Institute of Brussels - An NGO established in 1978 as a center for cultural and social development \citep{kib}.	
\end{itemize}

\subsubsection{Tools}
Some of the findings that are listed under DHuBase section could be considered as evidence for Digital Asset Management Systems and Online Catalog as well. For example, The Kurdish Digital Library of Kurdish Institute of Paris and the archive at Kurdish Heritage Institute. There is also a database that handles the recorded sounds for studying Kurdish dialects at the University of Manchester \citep{Kurdishdialects1}.

\subsubsection{Language}

We used the number of entires in Google search engine and other resources such as Wikipedia for the language \textit{visibility}. We also use the BLARK for Kurdish~\citep{hassani2018blark} for both the language \textit{computability} and its BLARK status.

\subsubsection{Digital Media}

The data for this section has been collected through searching the Internet and by consulting informed individuals. However, to quantify the wikis and blogs in Kurdish is not an easy task. The search engines such as Google support advanced features that allows user to select a target language. Unfortunately, Kurdish is currently not among the supported languages\footnote{This article was written before Google Translate added Kurmanji to its list. However, our understanding of a language is in considering it in its entirety. That is, a certain dialect of a language is resourceful while the other dialects are not, we do not assume the whole language to be resourceful.}. These means one should develop a web crawler to do the task or rely on normal search by looking for some random Kurdish texts of different Kurdish dialects and then to examine the results. This is a time consuming task with the result that might not be precise or reliable. However, it could be appropriate to test the visibility on the Internet. This is what followed in this research. Applying this approach, Google search engine yielded 331,000 and 351,1000 pages in average for Kurmanji and Sorani respectively, which can be considered as \textit{considerable}.  

\subsubsection{Education and Research}

We collected the data for this section through two different tools; one through an interview with scholars and the other using search engines and online databases. For the former, we conducted several interviews with scholars who are active in humanities studies and whose focus are the Kurds and Kurdish related issues. For the latter, we searched certain keywords and recorded the results. Below these tasks and the resulted data are presented.

The interviews were conducted based on some open-ended questions in order to explore the following areas:
\begin{itemize}
\item The role of digital resources in the research.
\item The level of familiarity of the scholars with the Digital Humanities concept.
\item The experienced obstacles and hindrances in front of research because Kurdish is not computable.
\end{itemize}

Table~\ref{tbl:DH-Intrv} shows a summary of the findings in the interviews.

\begin{table}[htbp]
\begin{center}
	\begin{tabular}{|p{3.7cm}|p{4.2cm}|p{2.2cm}|}
		\hline
		Subject & Responses to the subject & Respondents~\%\\ \hline
		Familiarity with Digital Humanities & Not at all & 100\% \\ \hline
		The role of digital resources & Important and very important & 70\% \\ \hline
		The role of digital resources & Not important at all or has a trivial impact & 30\% \\ \hline
		Obstacles because of lack of Language Technology & Not known for the scholars & 60\% \\ \hline
		Obstacles because of lack of Language Technology & Known for the scholars and assumed as an important factor & 40\%~ \\ \hline
	\end{tabular}
	\captionof{table}{Summary of findings in the interviews with the scholars} \label{tbl:DH-Intrv}
\end{center}
\end{table}

The online search was conducted with three aims in mind. Firstly, to find the available online resources related to DH in the context of Kurdish language. Secondly, to find evidence on Digital Humanities in the Iraqi Kurdistan Region's academic institutions. Finally, to compare the outcomes with the computationally rich languages such as English.

Table~\ref{tbl:se1} shows the results of the search on the Internet\footnote{Data was retrieved on during May 30 to June 14, 2015. Screenshots are available.}using keywords ``digital humanities + Kurdish language'' or ``digital humanities + Kurdish''\footnote{Some search engines such as SpringerLink, JSTOR, EEBSCOhost facilitate the search by letting users to indicate that all participated keywords must appear in a text. In this case the key word ``language'' would be redundant and would lead to more irrelevant results.}.

\begin{table}[htbp]
\begin{center}
	\begin{tabular}{|p{6cm}|p{1.5cm}|p{2.7cm}|}
		\hline
		Engine/Database & Results Found & Relevance\\ \hline
		Google & 12,400 & Partly\footnote{First 50 item investigated. Two items were found in which one could find an indirect correlation between the searched keywords.}\\ \hline
		Google Scholar & 1,550 & No relevance\footnote{First 50 item investigated. Only one found related. The document was about digital culture, in which in two occasions Kurdish is referred to.} \\ \hline
		SpringerLink & 2 & No relevance\footnote{The search without "language" resulted two items, neither of which was relevant.}\\ \hline
		Springer & 311 & No relevance\footnote{No evidence about Kurdish language} \\ \hline
		JStore & 85 & No relevance\footnote{None was related to Digital Humanities in the context of Kurdish language.}\\ \hline
		EBSCOhost-Academic Search Premier & 0 & No relevance \\ \hline
	\end{tabular}
	\captionof{table}{Search engines and databases result for Digital Humanities and KL} \label{tbl:se1}
\end{center}
\end{table}

Table \ref{tbl:se2} shows the results of the search on the Internet using keywords ``digital humanities''\footnote{Data was retrieved on October June 1, 2015. Screenshots are available.}.

\begin{table}[htbp]
\begin{center}
	\begin{tabular}{|p{6cm}|p{1.5cm}|p{2.7cm}|}
		\hline
		Engine/Database & Results Found & Relevance \\ \hline
		Google & 2,011,500 & High relevance \\ \hline
		Google Scholar & 20,400 & High relevance\footnote{100 items were investigated in different pages.} \\ \hline
		Springer & 692 & Relevance\footnote{The top 100 items showed fair relevance. 30 items were closely related to DH.} \\ \hline
		JStore & 443& High relevance\footnote{First 100 item investigated.} \\ \hline
		EBSCOhost-Academic Search Premier & 661 & High relevance\footnote{100 case was looked in detail.} \\ \hline
	\end{tabular}
	\captionof{table}{Search engines result for DH} \label{tbl:se2}
\end{center}
\end{table}

Also, the websites of the universities in the Iraqi Kurdistan Region were searched for ``digital humanities''. Table~\ref{tbl:su1} shows the result\footnote{Data was retrieved during June 8-10, 2015. Screenshots are available.}.

\begin{table}[htbp]
\begin{center}
	\begin{tabular}{|p{6.7cm}|p{1.5 cm}|p{2.4 cm}|}
		\hline
		University & Results Found & Relevance \\ \hline
		Salahaddin University-Erbil & 0 & - \\ \hline
		University of Sulaimani & 0 & - \\ \hline
		University of Kurdistan Hewl\^er & 0 & -\footnote{Website does not provide search facility. It was searched manually.} \\ \hline
		Koya University & 0 & - \\ \hline
		University of Duhok & 0 & - \\ \hline
		University of Zakho & 0 & - \\ \hline
		The American University of Iraq - Sulaimani & 1 & Relevance\footnote{The item that was found was about an academic member profile whose research interest included Digital Humanity and Digital Heritages.} \\ \hline
		American University Duhok Kurdistan & 0 & - \\ \hline
		Soran University & 0 & - \\ \hline
		Ishik University & 0 & - \\ \hline
	\end{tabular}
	\captionof{table}{Result for Digital Humanities search on the Universities' websites in the Iraqi Kurdistan Region} \label{tbl:su1}
\end{center}
\end{table}

To summarize, Table~\ref{tbl:se2} shows that DH has received considerable attention internationally. However, Table~\ref{tbl:se1} and Table~\ref{tbl:su1} show that there is no evidence about the activities related to Digital Humanities in the Kurdish language context, either internationally or regionally.

For this, we repeated the search, this time in two steps. The first step was to use other search keywords, which could represent a broader coverage of DH. The second step was to search specific websites based on a priori knowledge in the research area.
For the first approach the following combination of keywords were used:

\begin{enumerate}
\item text classification + Kurdish
\item information retrieval + Kurdish
\item digitization + Kurdish
\end{enumerate}

During this attempt, the above criteria were applied to the same search engines, which were used in the primary search. The results did not change the previous figures, therefore, the details are not repeated. This showed that despite using more inclusive interpretation of DH, it is still not a known subject in the context of Kurdish language.

The overall findings have been summarized in the Kurdish DHuRAF Table~\ref{tbl:dhu1}.

\begin{table}[htbp]
\begin{center}
\begin{tabular}{|l|*{3}{c|}|c|}
	\hline
	\cline{1-3}
	\multicolumn{1}{|l|}{\textbf{ Sections}}&\multicolumn{1}{c|}{\textbf{ Importance }}&\multicolumn{1}{c|}{\textbf{ Availability}} \\ 
	\cline{1-3}
	\hline
	\multicolumn{3}{|l|}{\textbf{ DHuBase}} \\
	\hline
	 Digitized Books& +++ & \\
	\hline
	 Digitized Photo Archives& &  \\
	\hline
	 Digitized Sound Archives& +++ & 1 \\
	\hline
	\multicolumn{3}{|l|}{\textbf{ Tools}} \\
	\hline
	 Digital Asset Management System& & 0 \\
	\hline
	 Online Catalog& & 0 \\
	\hline
	 Video and Film Analyzer& & 0 \\
	\hline
	\multicolumn{3}{|l|}{\textbf{ Language}} \\
	\hline
	 Visibility on the Internet& +++ & 2 \\
	\hline
	 Computability& +++ & 0 \\
	\hline
	 BLARK Status& +++ & 1 \\
	\hline
	
	\multicolumn{3}{|l|}{\textbf{ Digital Medeia}} \\
	\hline
	 News Agency& +++ & 2 \\
	\hline
	 News Agency Website & +++ & 2 \\
	\hline
	 Satellite TV & ++ & 2 \\
	\hline
	 Satellite TV Website & ++ & 1 \\
	\hline
	 Local TV & ++ & 2 \\
	\hline
	 Social Media (Twitter, Facebook, ...) & ++ & \\
	\hline
	 Blog& ++ & 1 \\
	\hline
	 Wiki& ++ & 1 \\
	\hline
	\multicolumn{3}{|l|}{\textbf{ Education}} \\
	\hline
	 Academic Awareness& +++ & 0 \\
	\hline
	 Active Institution& ++ & 0 \\
	\hline
	 Academic Program-UG & & 0 \\
	\hline
	 Academic Program-Master& & 0 \\
	\hline
	 Academic Program-PhD& & 0 \\
	\hline
	\multicolumn{3}{|l|}{\textbf{ Research}} \\
	\hline
	 Project-Finished& + & 0 \\
	\hline
	 Project-Ongoing& ++ & 0 \\
	\hline
	 Project-Canceled& +++ & 0 \\
	\hline
	 Cumulative Fund& +++ & 0 \\
	\hline
\end{tabular}
\caption[Caption title in LOF]{DHuRAF Indicators - Kurdish Case} \label{tbl:dhu1}
\end{center}
\end{table}

\section{Applying DHuRAF: Scottish Gaelic}

The second case to investigate the efficiency of DHuRAF was Scottish Gaelic. The following sections provide the results of applying DHuRAF and the analysis of the results.

\subsection{Findings}

The findings have been documented according to the DHuRAF sections and have been summarized using the DHuRAF Indicators. This section explains how the findings have been obtained and shows the DHuRAF Indicators table.  

\subsubsection{DHuBase}

For DHuBase, used the Internet with ``Scottish Gaelic digital resource''.

\begin{itemize}
\item Digital Archive of Scottish Gaelic - It ``aims to provide a comprehensive electronic corpus of Scottish Gaelic texts for students and researchers of Scottish Gaelic language, literature and culture''~\citep{ScottishGaelic1}. Currently, tt includes three main sections which are corpus, fieldwork, and jukebox.
\item Gaelic Resources - Primarily, A group for  
\item Growing Gaelic - Aiming at improving the Gaelic status in Scotland~\citep{ScottishGaelic3}.
\item Am Faclair Beag - Online Gaelic-English dictionary~\citep{ScottishGaelic4}.   
\item St\`or-d\`ata - Online Gaelic-English dictionary~\citep{ScottishGaelic5}.   
\item Thesaurus - Online Gaelic-English thesaurus~\citep{ScottishGaelic6}.   
\item Scottish Gaelic dictionary - Despite the name, this page includes links to different Gaelic dictionaries, persons and places names, language and literature resource ~\citep{ScottishGaelic7}.   
\item Tobar an Dulachais - ``This website contains over 38,000 oral recordings made in Scotland and further afield, from the 1930s onwards.''~ ~\citep{ScottishGaelic8}.   
\item ambile-highland history \& culture - Includes ``photographs, illustrations, rare books and documents, as well as short films, audio recordings, interactive games and comics.''~ ~\citep{ScottishGaelic9}.   
\item Hebridean Connections - Over 40,000 records about history, traditions, culture,  archaeology, and the genealogy of Scots~\citep{ScottishGaelic10}.   
\end{itemize}

\subsubsection{Tools}
\begin{itemize}
	\item Online catalog the Gaelic learning materials - In this online collection, the library of the University of the Highlands and Islands provide a section list of the Gaelic learning materials~\citep{ScottishGaelic11}.
\end{itemize}

\subsubsection{Language}

Gaelic relatively lacks of resources\citep{lamb2016developing}. Importantly, we could not find an accounting status such as BLARK to let us to label the status of the language with regard to NLP. However, this can be mentioned that several projects have been conducted by a small group of active scholars such as Kevin Patrick Scannell(see~\cite{scannelllinks} and Annotated Reference Corpus of Scottish Gaelic (ARCOSG) by \cite{arbuthnot2016annotated}).

\subsubsection{Digital Media}

The data for this section has been collected through searching the Internet. We used Google as the search engine. Although Gaelic is currently not among the supported languages, Manx and Gaeilge (Irish Gaelic) are supported int the advanced search. We based our search on various Gaelic sentences in order to check the visibility on the Internet. We build our sentences\footnote{For example, ``anns bhith g\`aidhlig airson tha agus''.} based on a frequency list\footnote{http://www.foramnagaidhlig.net/foram/viewtopic.php?t=2508}. This is what followed in this research. Applying this approach, Google search engine yielded 2,220,000 results, which can be considered as \textit{vast}.  

\subsubsection{Education and Research}

We collected the data for this section using search engines and online databases based on certain keywords. The results are shown below.

We conducted with three aims in mind. Firstly, to find the available online resources related to DH in the context of Gaelic language. Secondly, to find evidence on Digital Humanities in the academic institutions with regard to Gaelic studies, focusing on the institutions located in United Kingdom (particularly, Scotland and Ireland). Finally, to compare the outcomes with the computationally rich languages such as English.

Table~\ref{tbl:se3} shows the results of the search on the Internet\footnote{Data was retrieved on during July 1 to June 18, 2017. Screenshots are available.} using keywords ``digital humanities + Gaelic language'' or ``digital humanities + Gaelic''\footnote{Some search engines such as SpringerLink, JSTOR, EEBSCOhost facilitate the search by letting users to indicate that all participated keywords must appear in a text. In this case the key word ``language'' would be redundant and would lead to more irrelevant results.}.

\begin{table}[htbp]
\begin{center}
	\begin{tabular}{|p{6cm}|p{1.5cm}|p{2.7cm}|}
		\hline
		Engine/Database & Results Found & Relevance\\ \hline
		Google & 55,600 & High\footnote{First 50 item investigated.} \\ \hline
		Google Scholar & 2,320 & Partly relevant\footnote{First 50 item investigated.} \\ \hline
		SpringerLink & 0 & NA \\ \hline
		Springer & 0 & NA  \\ \hline
		JStore & 195 & High\footnote{None was related to Digital Humanities in the context of Kurdish language.}\\ \hline
		EBSCOhost-Academic Search Premier & 29 & High \\ \hline
	\end{tabular}
	\captionof{table}{Search engines and databases result for Digital Humanities and Gaelic} \label{tbl:se3}
\end{center}
\end{table}

Table~\ref{tbl:se4} shows the results of the search on the Internet using keywords ``digital humanities''\footnote{Data was retrieved on October June 1, 2015. Screenshots are available.}.

\begin{table}[htbp]
\begin{center}
	\begin{tabular}{|p{6cm}|p{1.5cm}|p{2.7cm}|}
		\hline
		Engine/Database & Results Found & Relevance \\ \hline
		Google & 2,011,500 & High relevance \\ \hline
		Google Scholar & 20,400 & High relevance\footnote{100 items were investigated in different pages.} \\ \hline
		Springer & 692 & Relevance\footnote{The top 100 items showed fair relevance. 30 items were closely related to DH.} \\ \hline
		JStore & 443& High relevance\footnote{First 100 item investigated.} \\ \hline
		EBSCOhost-Academic Search Premier & 661 & High relevance\footnote{100 case was looked in detail.} \\ \hline
	\end{tabular}
	\captionof{table}{Search engines result for DH} \label{tbl:se4}
\end{center}
\end{table}

Also, the websites of the universities in the United Kingdom were searched for ``digital humanities''. Table~\ref{tbl:su2} shows the result\footnote{Data was retrieved during July 17-18, 2017. Screenshots are available.}.

\begin{table}[htbp]
\begin{center}
	\begin{tabular}{|p{6.7cm}|p{1.5 cm}|p{2.4 cm}|}
		\hline
		University & Results Found & Relevance \\ \hline
		University of Aberdeen & 285 & High \\ \hline
		Robert Gordon University & 108 & High \\ \hline
		University of Glasgow & 158 & High \\ \hline
		University of Edinburgh & 2,320 & High \\ \hline
		University of St Andrews & 973 & High \\ \hline
		University of Dublin & 5 & High \\ \hline
		Trinity College Dublin & 1630 & High  \\ \hline
	\end{tabular}
	\captionof{table}{Result for Digital Humanities search on the Universities' websites in Scotland and Ireland} \label{tbl:su2}
\end{center}
\end{table}

To summarize, Table \ref{tbl:se4} shows that DH has received considerable attention internationally. Also, Table~\ref{tbl:se3} and Table~\ref{tbl:su2} show that there are highly considerable activities related to Digital Humanities in the two regions/countries (Scotland and Ireland) of the UK where Gaelic is considered a native language.

The overall findings have been summarized in the Gaelic DHuRAF Table \ref{tbl:dhu2}.

\begin{table}[htbp]
\begin{center}
\begin{tabular}{|l|*{3}{c|}|c|}
\hline
\cline{1-3}
\multicolumn{1}{|l|}{\textbf{ Sections}}&\multicolumn{1}{c|}{\textbf{ Importance }}&\multicolumn{1}{c|}{\textbf{ Availability}} \\ 
\cline{1-3}
\hline
\multicolumn{3}{|l|}{\textbf{ DHuBase}} \\
\hline
Digitized Books& +++ & 3 \\
\hline
 Digitized Photo Archives& + & 2 \\
\hline
 Digitized Sound Archives& +++ & 3 \\
\hline
\multicolumn{3}{|l|}{\textbf{ Tools}} \\
\hline
 Digital Asset Management System& & 2 \\
\hline
 Online Catalog& & 2 \\
\hline
 Video and Film Analyzer& & 3 \\
\hline
\multicolumn{3}{|l|}{\textbf{ Language}} \\
\hline
 Visibility on the Internet& +++ & 3 \\
\hline
 Computability& +++ & 3 \\
\hline
 BLARK Status& +++ & NA \\
\hline

\multicolumn{3}{|l|}{\textbf{ Digital Medeia}} \\
\hline
 News Agency& +++ & 4 \\
\hline
 News Agency Website & +++ & 4 \\
\hline
 Satellite TV & ++ & 2 \\
\hline
 Satellite TV Website & ++ & 1 \\
\hline
 Local TV & ++ & 2 \\
\hline
 Social Media (Twitter, Facebook, ...) & ++ & \\
\hline
 Blog& ++ & 1 \\
\hline
 Wiki& ++ & 1 \\
\hline
\multicolumn{3}{|l|}{\textbf{ Education}} \\
\hline
 Academic Awareness& +++ & 4 \\
\hline
 Active Institution& ++ & 10+ \\
\hline
 Academic Program-UG & & 0 \\
\hline
 Academic Program-Master& & 0 \\
\hline
 Academic Program-PhD& & 0 \\
\hline
\multicolumn{3}{|l|}{\textbf{ Research}} \\
\hline
 Project-Finished& + & 0 \\
\hline
 Project-Ongoing& ++ & 0 \\
\hline
 Project-Canceled& +++ & 0 \\
\hline
 Cumulative Fund& +++ & 0 \\
\hline
\end{tabular}
\caption[Caption title in LOF]{DHuRAF Indicators - Scottish Gaelic Case} \label{tbl:dhu2}
\end{center}
\end{table}

\section{Discussion}
\label{sec:discussion}
The research aimed to answer this question: ``how can we assess the state of research in DH in minority languages, resource-scarce languages, and languages with dialect diversity?''
 
To answer the question, we suggested a framework which guides us in presenting a of the situation by which we can infer the status of DH in the specific context. The framework then was used in the context of Kurdish and Gaelic languages. These language were chosen as samples of the languages with the characteristics mentioned in the above question. The result of applying it were shown in Table~\ref{tbl:dhu1} and Table\pageref{tbl:dhu2} for Kurdish and Gaelic respectively.

The result shows that DH in the Kurdish studies context can be labeled as \textit{Void}, which as the framework defines it, suggests that DH as a discipline, currently does not exist in the mentioned context. However, there are evidence of using the technology in Kurdish related studies, though these are not characterized as DH studies. Nevertheless, the results show that the usage of information technology for Kurdish related studies is emerging. However, the extent, depth, organization, and structure of these activities need more advancement, if one compares them to equivalent resources for languages with well-established DH studies.

To illustrate, this comparison can be performed with a language such as English, as a global language, or Swedish, which is spoken by a population with the magnitude of 1/3 of people who speak Kurdish. In the former case a simple search leads to millions of resources, just in the surface of the Internet and not by searching deep in subscription required resources. The latter case makes the situation statistically even more interesting. That is, only one of the explored resources, the National Library of Sweden, includes almost 8 million hours of audio and moving image recordings~\citep{nationalLibSweden}. As another example, the Swedish Language Bank, which is managed by the University of Gothenburg, has been collecting a corpora for Swedish language since 1975, which currently exceeds one million records~\citep{swedishLangBank}.

However, for the Gaelic case, the result shows that Gaelic DH is in its~\textit{Infancy}. This, as the framework defines it, suggests that there are indications of existence of the basic apparatus for DH studies in Gaelic, but there is no evidence that suggests a significant DH study in Gaelic.  

To summarize, the findings showed that there are supporting evidence indicating the activities which could be considered as part of Digital Humanities related tasks in the Kurdish context. However, researchers would not be able to utilize these resources in the absence of Kurdish language technology. Moreover, the quantity and diversity of theses resource are still very limited in comparison with the popular languages such as English or even not popular such as Swedish. However, the Gaelic situation is different, that is, although the Gaelic DH is in its infancy, it has the potential to be expanded and leveraged to an active status. 

\subsubsection{Web as a language resource}

\cite{boydcritical2012}\footnote{In writing danah boyd's name, we faced a dilemma, which was whether to write her name as she mentioned in {http://www.danah.org/name.html} or capitalize it as the format of this article and its bibliographical norms required. With all due respect to danah's decision and approach, we preferred to stay conservative and to follow the traditional approach. However, this is somehow ironic, because in Kurdish language, usually, there is no capitalization form for names. In fact, in Persian/Arabic script there is no proper way to express it, and in Latin script it has not generally been practiced. We have studied about naming and proper nouns in Kurdish with regard to computational linguistics and the issues that it makes.} have critically discussed the role of big data in different studies.

Particularly,~\cite{kilgarriffintroduction2003} have presented the issues that should be of concerns when one uses web as corpus. However, in the context of Kurdish language, web contents can serve as an important resource, where the big data is missing. It can be used as a bridge over the digitally less-resourced situation of which the Kurdish context is currently suffering, and the situation that the projects for making the language computationally rigorous. The key point here is that this approach is also needs to be used with vigilance, because even on the web, Kurdish is not resource-rich and the available contents are biased in different ways~\citep{candan2008nation}.

\section{Conclusion}
Despite being a relatively new subject, the attention to Digital Humanities is steadily growing in academia. However, the level of the attention to the subject and its applicability in different contexts are highly related to the readiness and maturity of the required foundation. One of the major factors is that the target context is computationally ready for the research. To assess the DH readiness we developed an accounting framework. This framework, DHuRAF, consists of 5 sections, which together they provide an overview of the DH status in a particular context. As an experiment, the framework was applied to explore the situation of Digital Humanities in the context of Kurdish and Gaelic communities. The result showed that the prerequisites are not ready for the active DH studies and research in the Kurdish context while they are at a reasonable level to let the Gaelic DH studies to foster. As for the Kurdish case, the research concluded that the its DH situation could be, at least partly, because of the language is not yet computable, while in the Gaelic case, not only the language is supported by more computational facilities but also a broader set of digital artifacts have made the environment more appropriate in order to allow the interested researchers to embark in Gaelic DH studies.

Our experiment focused on the contexts of minority languages, therefore, further studies are required in order to show whether DHuRAF is applicable to other contexts regardless of the dominance of particular languages. Also the result suggests that in particular cases such as Kurdish, DH might be able help NLP by to provide a synergy to overcome the current obstacles that exist in front of both DH and NLP studies . That is, lack of resources to support NLP can be compensated by utilizing available DH resources, which in return it can nurture back the DH by improving the NLP capacity. Further works in the future on applying DHuRAF on other languages may suggest revised versions with enhancement/improvements in the sections, items, and the evaluation scheme. All together, these could bring DH and NLP more closer and make them more influential on each other. 

\begingroup
\theendnotes
\endgroup

\section*{Acknowledgement}

The authors would like to appreciate Kirstin Crawford for reading the early version of this article. They would also like to appreciate the feedback from Professor Daniel Paul O'Donnell from the journal of ``Digital Studies / Le champ num\'erique'', Dr. Jonathan D. Fitzgerald and Gregory J. Palermo from the journal of ``digital humanities quarterly'', and also the anonymous reviewers from the both journals who have commented on this paper.

\bibliographystyle{chicago}
\bibliography{DHuRAF}

\end{document}